\documentclass{epl}

\title{A promising method for the measurement of the local acceleration of gravity using Bloch oscillations of ultracold atoms in a vertical standing wave}
\author{Pierre~Clad\'e\inst{1}, Sa\"\i
da~Guellati-Kh\'elifa\inst{2}, Catherine~Schwob\inst{1}, Fran\c
cois~Nez\inst{1}, Lucile~Julien\inst{1} and Fran\c
cois~Biraben\inst{1}} \institute{ \inst{1} Laboratoire Kastler
Brossel, \'Ecole Normale Sup\'erieure, CNRS, UPMC, 4 place
Jussieu, 75252 Paris Cedex 05, France\\
\inst{2}  CNAM-INM, Conservatoire National des Arts et M\'etiers,
292 rue Saint Martin, 75141 Paris Cedex 03, France}

\pacs{32.80.Pj}{Optical cooling of atoms; trapping}
\pacs{42.50.Vk}{Mechanical effects of light on atoms, molecules,
electrons, and ions} \pacs{04.80.-y}{Experimental studies of
gravity}

\shortauthor{P. Clad\'e \etal}

\shorttitle{Measurement of the local acceleration of gravity using
Bloch oscillations}

\begin{document}
\maketitle

\begin{abstract}
An obvious determination of the acceleration of gravity $g$ can be
deduced from the measurement of the velocity of falling atoms
using a $\pi-\pi$ pulses sequence of stimulated Raman transitions.
By using a vertical standing wave to hold atoms against gravity,
we expect to improve the relative accuracy by increasing the
upholding time in the gravity field and to minimize the systematic
errors induced by inhomogeneous fields, owing to the very small
spatial amplitude of the atomic center-of-mass wavepacket periodic
motion. We also propose to use such an experimental setup nearby a
Watt balance. By exploiting the $g/h$ ($h$ is the Planck constant)
dependence of the Bloch frequency, this should provide a way to
link a macroscopic mass to an atomic mass.
\end{abstract}

\section{Introduction}
The dynamics of an atomic wave packet in a periodic potential
under the influence of a static force has been extensively
analyzed using different physical approaches : in terms of
Wannier-Stark resonance states \cite{Wilkinson}, Bloch
oscillations \cite{Dahan} or macroscopic quantum interferences
induced by tunnelling due to the external acceleration
\cite{Anderson}(for review see \cite{Raizen}). An interesting
configuration occurs when the external force is induced by the
acceleration of gravity. In this case the Bloch frequency is equal
to $\nu_B = \frac{m g \lambda}{2 h}$ and depends only on the local
acceleration of the gravity $g$, the wavelength of the light
$\lambda$ and some fundamental constants. This frequency is
typically in the range $100~Hz-2000~Hz$, and its measurement
allows the determination of $g$. Previous experiments have already
been realized using the dynamics of BEC \cite{Anderson,Roati} or
degenerated Fermi gas \cite{Roati} in vertical 1-D optical
lattice. Kasevich's group has observed the interference between
macroscopic quantum states of BEC atoms confined in a vertical
array of optical traps. This interference arises from the
tunneling induced by the acceleration of gravity and appears as a
train of falling atomic pulses. The acceleration of the gravity
$g$ was determined by measuring the spatial period of the pulses
train. In Ingusio's group \cite{Roati} the Bloch period is
straightforwardly deduced from the evolution of the momentum in
the trap by adiabatically releasing the cloud from the lattice. In
both experiments the detection is performed by imaging the falling
atomic cloud using absorption imaging techniques and the
uncertainty on the $g$ determination did not exceed $10^{-4}$
dominated by the imaging system.

\begin{figure}
\onefigure[scale=0.5]{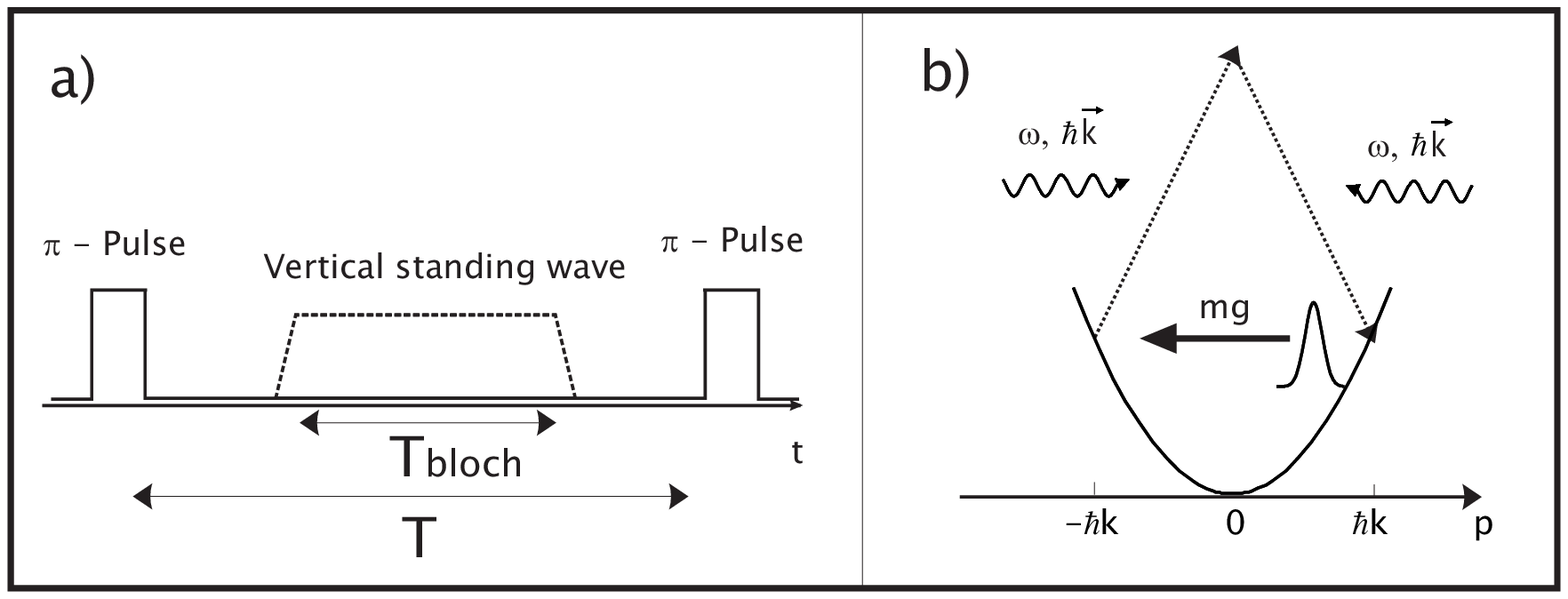} \caption{a). Experimental
pulses sequence. b) A narrow velocity class is selected by the
first Raman pulse. When the vertical standing wave is switched on,
atoms fall until they are resonant with the $\Lambda$ transition:
they absorb a photon from the upward wave and emit a stimulated
photon in the downward wave. This induces a momentum exchange of
$2 \hbar k$.} \label{f.1}
\end{figure}

Our experimental approach is based on the precise determination of
the velocity distribution of atoms along the vertical axis using
Doppler-sensitive Raman transitions. An obvious determination of
the acceleration of the gravity is possible by measuring the
atomic velocity variation after a given falling time $T$. This is
performed by applying a $\pi-\pi$ pulses sequence with a spacing
time $T$: the first pulse defines an initial velocity by selecting
a narrow velocity class from an ultracold atomic sample. Atoms are
then in a well defined internal state. The second pulse measures
the final velocity distribution of the atoms after the fall by
transferring a resonant velocity slice to another internal state.
This so-called velocity sensor allows us to locate the center of
the velocity distribution with high accuracy and is now limited by
the experimental setup platform's vibrations \cite{clade}. We
could substantially improve  the relative uncertainty on the
measurement of $g$ by increasing the falling time $T$, but this
parameter is swiftly limited by the dimension of the vacuum
chamber. In our experiment we suggest to hold the atoms against
gravity by applying between the two Raman pulses a far resonant
standing wave, during an interrogation time $T_{Bloch}$
(fig.~\ref{f.1}.a). In the non dissipative case, atoms fall until
they absorb a photon from the upward wave and emit a stimulated
photon in the downward wave. The atoms make a succession of
$\Lambda$ transitions (fig.~\ref{f.1}.b) inducing a momentum
exchange of $2\hbar k$ ($k$ is a wave vector of the standing wave)
with the cycling frequency $\nu_B = \frac{m g \lambda}{2h}$. This
evolution is equivalent to the dynamics of the Bloch oscillations
\cite{Peik}. In a previous work \cite{Battesti} we have measured a
transfer efficiency of $99.5\%$ per cycle. This result promises a
large number of Bloch oscillations. Another particular interest of
our method is the small spatial amplitude of the atomic wavepacket
motion, during the oscillations \footnote{The amplitude of the
oscillation of atomic wavepacket is given by $\Delta z =
\Delta_n/2|F|$, where $\Delta_n$ is the energy width of the nth
band \cite{Dahan}. In our experiment only the fundamental band is
considered $\Delta_0 < E_r$ , $F = mg$ Then $ \Delta z < 1\mu m $
which is very small compared to the size of the atomic cloud
($1~mm$).}, therefore the selection (first Raman pulse) and
measurement (second Raman pulse) are done in a small volume
allowing a better control of systematic effects arising from
inhomogeneous fields. In this letter we investigate the
possibility to make a high precise measurement of the acceleration
of gravity $g$ using Bloch oscillations of cold atoms in a
vertical standing wave. Such accurate measurements have important
repercussions on geophysical applications including earthquake
predictions, locating oil and studies of the global warming. We
first describe our measurement method, then we present a
preliminary measurement of $g$, and finally we discuss the signal
losses observed when we increase the number of Bloch oscillations.
\section{Experimental Set-up}

The main experimental apparatus has already been described in
reference \cite{Battesti}. Briefly, $^{87}Rb$ atoms are captured,
from a background vapor, in a $\sigma^+-\sigma^-$ configuration
magneto-optical trap (MOT). The trapping magnetic field is
switched off and the atoms are cooled to about $3~\mu K$ in an
optical molasses. After the cooling process, we apply a bias field
of $\sim100~\mathrm{mG}$. The atoms are then optically pumped into
the $F=2, m_F=0$ ground state. The determination of the velocity
distribution is performed using a $\pi-\pi$ pulses sequence of two
vertical counter-propagating laser beams (Raman beams): the first
pulse with a fixed frequency $\nu_{sel}$, transfers atoms from $5
S_{1/2}$, $\left|F=2, m_F = 0\right>$ state to $5 S_{1/2}$,
$\left|F=1,m_F = 0\right>$ state, into a velocity class of about
$v_r/15$ centered around $(\lambda \nu_{select}/ {2}) - v_r$ where
$\lambda$ is the laser wavelength and $v_r$ is the recoil
velocity. To push away the atoms remaining in the ground state
F=2, we apply after the first $\pi$-pulse, a laser beam resonant
with the $5S_{1/2}~(F=2)$ to $5P_{1/2}~(F=3)$ cycling transition.
Atoms in the state F=1 fall under the acceleration of gravity
during $T$. We then perform the final velocity measurement using
the second Raman $\pi$-pulse, whose frequency is $\nu_{meas}$. The
population transfer from the hyperfine state $F=1$ to the
hyperfine state $F=2$ due to the second Raman pulse is maximal
when $2\pi(\nu_{sel}-\nu_{meas}) = g\times T\times
\parallel(\mathbf{k_1}-\mathbf{k_2})\parallel$, where
$\mathbf{k_1}$, $\mathbf{k_2}$  are the wave vectors of the Raman
beams. The populations ($F=1$ and $F=2$) are measured separately
by using the one-dimensional time of flight technique developed
for atomic clocks and depicted in \cite{Clairon}. To plot the
final velocity distribution we repeat this procedure by scanning
the Raman beam frequency $\nu_{meas}$ of the second pulse.

\begin{figure}
\twofigures[scale=0.6]{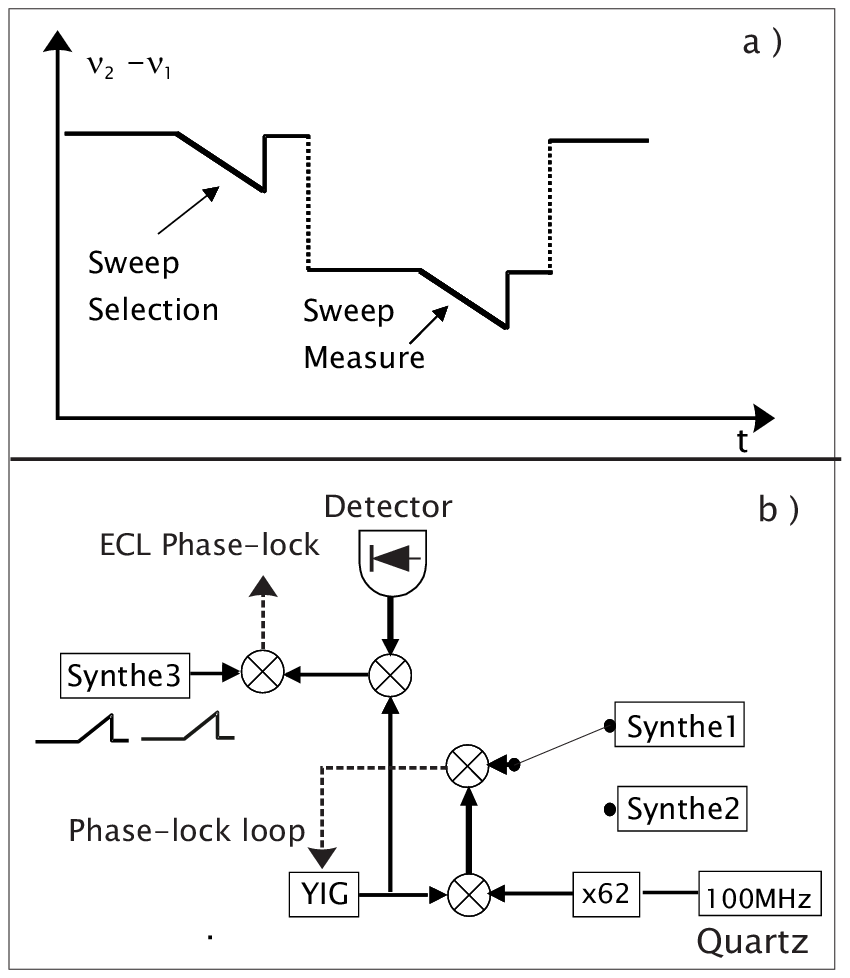}{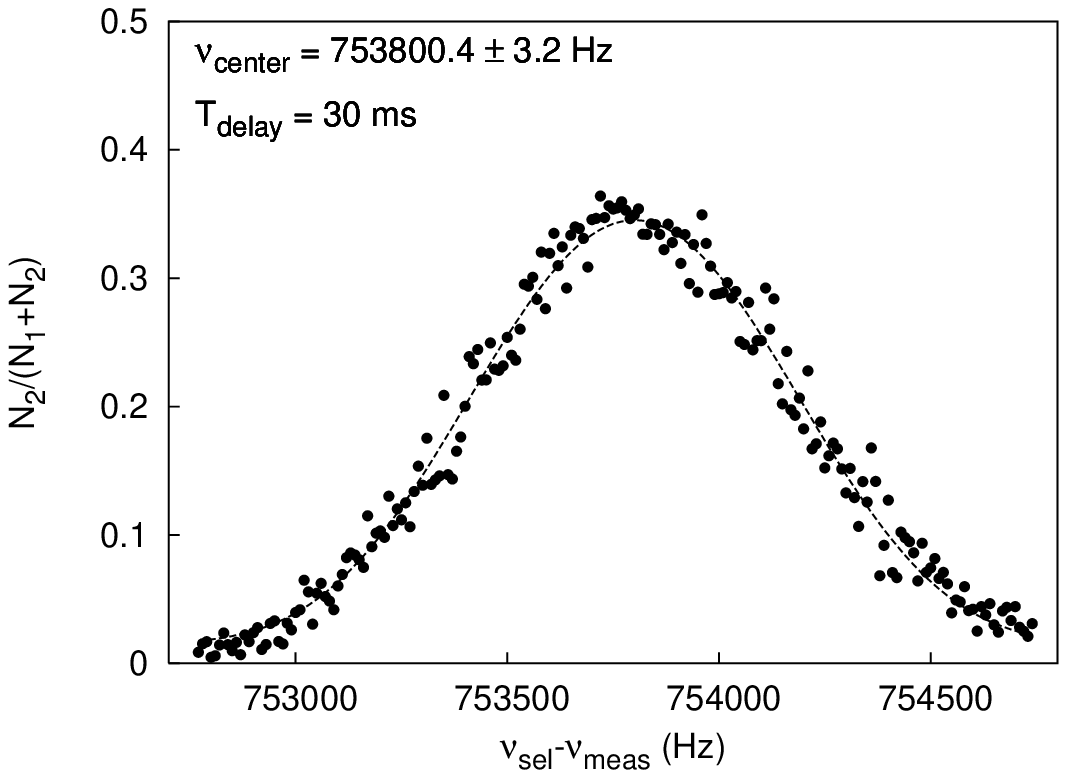} \caption{ a).
Temporal variation of the frequency difference of the two Raman
beams. b). The synthe1 and the synthe2 allow us to switch between
the selection and a measurement steps. To compensate the Doppler
shift during the falling of the atoms the frequency of the
synthes3 is swept linearly during the Raman pulses.} \label{f.2}
\caption{The final distribution velocity of atoms after a free
fall of $30~ms$. The center of the velocity distribution is
located with an uncertainty of $3.2~Hz$, corresponding to a
relative uncertainty of $4\times 10^{-6}$ on the measurement of
the acceleration of the gravity .} \label{f.3}
\end{figure}

The two Raman beams are generated using two diode lasers injected
by two extended-cavity diode lasers (ECLs). To drive the
velocity-sensitive Raman transition, the frequency difference  of
the master lasers must be precisely resonant with $^{87}Rb$
ground-state hyperfine transition ($\sim 6.8~\mathrm{GHz}$). The
frequency of one ECL is stabilized on a high stable Zerodur
Fabry-Perot (ZFP) cavity. The second ECL is then phase-locked to
the other using the beat note technique (see fig.~\ref{f.2}.b).
The very stable RF source is performed by mixing the $62th$
harmonic of a $100~\mathrm{MHZ}$ quartz oscillator with different
digital synthesizers (SRS DS345). They are used to tune finely the
frequency of the RF source. A YIG oscillator is phase-locked on to
the central line of the source in order to reject completely the
residual sidebands of the different mixings. A multiplexer
switches between two synthesizers to generate the frequency offset
for the velocity selection or measurement. The frequency of a
third synthesizer is linearly swept during the selection and the
measurement pulses to compensate the Doppler shift during the fall
of the atoms (fig.~\ref{f.2}.a). The time interval between the two
Raman pulses is precisely defined by the delay between the
triggering signal of the two frequency sweeps. Each laser beam
passes through an acousto-optic modulator ($\sim 80~\mathrm{MHz}$)
for timing (switch on and off) and intensity control.
The two beams have linear orthogonal polarizations and are coupled
into the same polarization maintaining optical fiber. The pair of
Raman beams is sent through the vacuum cell. The counter
propagating configuration is achieved using a polarizing
beam-splitter cube and an horizontal retroreflection mirror placed
above the exit window of the cell. The standing wave used to
create the 1-D optical lattice is generated by a Ti:Sapphire
laser, whose frequency is stabilized on the same highly stable ZFP
cavity. This laser beam is splitted in two parts. To perform the
timing sequence, each one passes through an acousto-optic
modulator to control its intensity and frequency. The beams are
detuned by 260~GHz from the $5S_{1/2}-5P_{3/2}$ resonance line to
avoid spontaneous emission. With these laser parameters, the
optical potential depth $U_0$ equals to $2.7~E_R$
($E_R=\frac{\hbar^2 k^2}{2m}$ is the recoil energy). For this
value, when the external acceleration is due to gravity, the
transfer of atoms to the higher bands remains insignificant for
several periods of Bloch oscillations.

\section{Results}
In a first experiment, we determine $g$ by measuring the atomic
velocity variation after the free fall of atoms during $30~ms$,
using the $\pi-\pi$ Raman pulses sequence described previously.
The typical final velocity distribution is shown in
(fig.~\ref{f.3}). The center of this distribution is located with
an uncertainty of $3.2~\mathrm{Hz}$ (corresponding to $v_r/5000$)
in an average time of $20~mn$, allowing a measurement of $g$ with
a relative uncertainty of $4\times 10^{-6}$. This uncertainty is
limited by many systematic errors. These errors may occur due to
the vibration noise of the retroreflecting mirror \cite{clade},
the fluctuation of the number of detected atoms and the atomic
motion between the two pulses (effect of inhomogeneous fields).

In a second experiment, we apply between the two $\pi$-pulses a
standing wave during an interrogation time $T_{Bloch}$. We then
study the evolution of the final momentum distribution by changing
$T_{Bloch}$. Before analyzing the experimental results, we briefly
recall the relevant results of Bloch's theory. The energy spectrum
of the particle presents a band structure (indexed by $n$) arising
from the periodicity of the potential (optical lattice with period
$d=\lambda/2$). The corresponding eigenenergies $E_n(q)$ and the
eigenstates $|n,q\rangle$ (Bloch states) are periodic functions of
the continuous quasi-momentum $q$, with a period $2k=2\pi/d$. The
quasi-momentum q is conventionally restricted to the first
Brillouin zone $]-\pi/d, \pi/d]$. If we apply a constant force
$F$, sufficiently weak in order to avoid interband transitions, a
given Bloch state $|n,q(0)\rangle$ evolves (up to a phase factor)
into the state $|n,q (t)\rangle$ according to

\begin{equation}
\label{e.1} q(t)=q(0) + 2 k \frac {t}{\tau_B} \pmod{2\pi/d}
\end{equation}

\begin{figure}
\twofigures[scale=0.8]{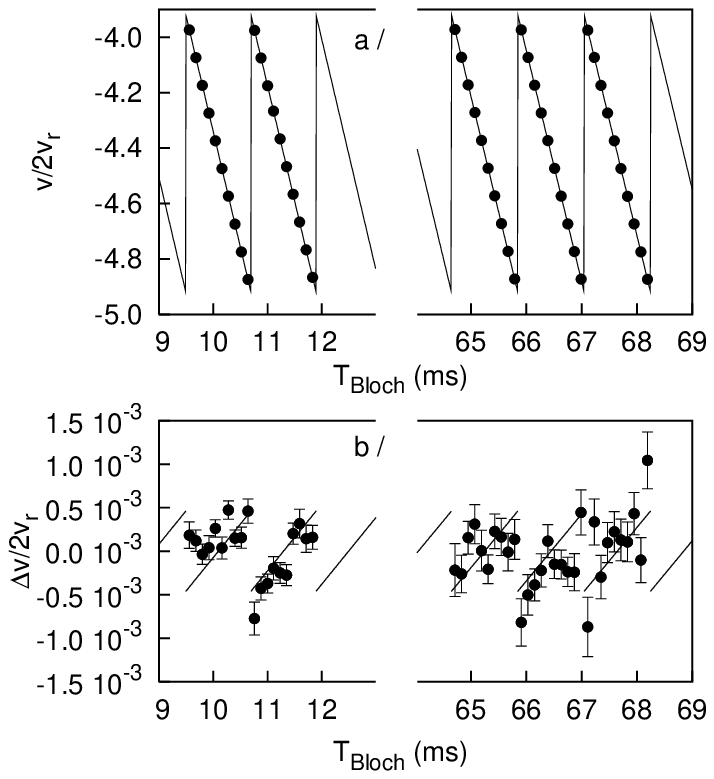}{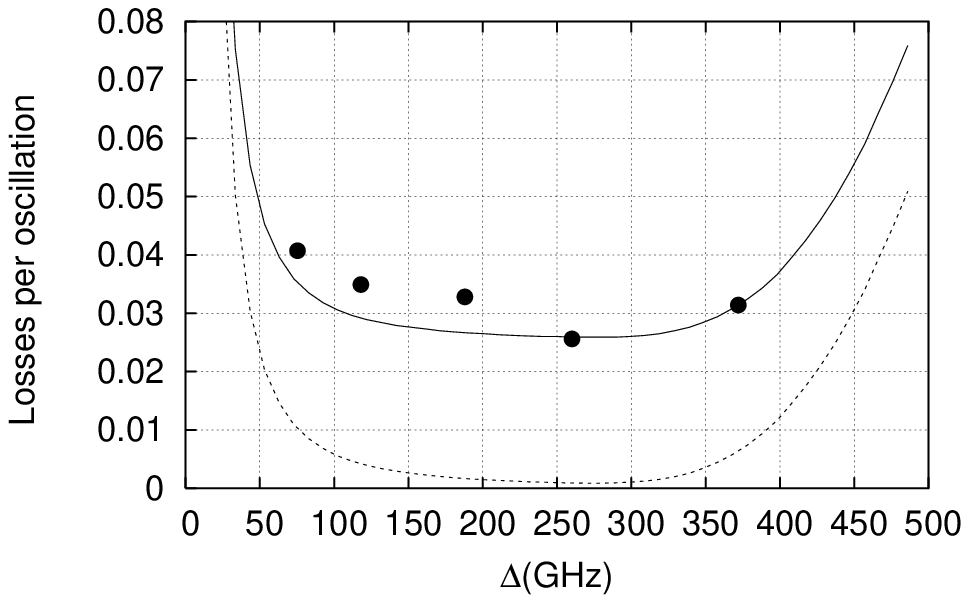}\caption{a. The
center of the final velocity distribution versus the duration of
the standing wave. The dot represent the experimental data and the
line the least-square fit performed by fixing the recoil velocity.
b) The residuals of the fit.} \label{f.4} \caption{The losses per
oscillation versus detuning $\Delta$. These losses are obtained by
comparing the number of atoms measured after $N$ oscillations to
those measured after $N=10$ oscillations. The relevant advantage
of this presentation is to take into account only the losses
during the Bloch oscillations process. Dot: the experimental data.
The theory: (Dotted line). Solid line: theory including the
interparticle collisions.} \label{f.5}
\end{figure}

When the atoms are only submitted to the gravity force, the Bloch
period $\tau_B$ is given by $\frac{2 h}{m g\lambda}$. This period
corresponds to the time required for the quasi-momentum to scan a
full Brillouin zone. In our experiment, first we prepare Bloch
states around $q=0$ (in lattice frame) at the bottom of the
fundamental energy band $(n=0)$ by turning on adiabatically the
standing wave (rise time of $300~\mu s$): this avoids a transfer
of population into the higher energy band. We point out that just
before turning on the Bloch potential, selected atoms reach a mean
velocity of about $10~v_r$. In order to compensate this velocity
drift, the upward beam's frequency is shifted by $\sim
150~\mathrm{kHz}$ (in the laboratory frame the standing wave is
then moving with a constant velocity of about $10~v_r$). To use a
pure standing wave, we should launch atoms in ballistic
atomic-fountain trajectories either from a moving molasses
\cite{ClaironII} or with Bloch oscillations \cite{Battesti}, and
turn on the Bloch potential when they reach their summit. After
time $T_B$ we suddenly switch off the optical potential and we
measure the final momentum distribution in the first Brillouin
zone. In (fig.~\ref{f.4}.a) we report the center of the measured
peak as a function of the holding time $T_B$. The observed
sawtooth shape is the signature of Bloch oscillations
(eq.~\ref{e.1}).  We observe more than $60$ Bloch periods
corresponding to the longest lived Bloch oscillator observed in
bosonic systems. To determine the Bloch period $\tau_B$, we
measure the time interval between the centers of the two extreme
slopes of sawtooth. We extract the value of $\tau_B$ by dividing
this time interval by the number of periods. This measurement
leads to a determination of the local acceleration of gravity with
a relative uncertainty of $1.1~10^{-6}$ as $\frac{h}{m}$ ratio
\cite{Mohr} and the wavelength $\lambda$ are known with a better
accuracy. The linear fit of the experimental data in
(fig.~\ref{f.4}.a) is performed by fixing the value of the recoil
velocity. In fig.~\ref{f.4}.b we present the residuals of the fit;
they increase when we move away from the center of the Brillouin
zone. That reveals that there is a difference between the
quasi-momentum and the measured momentum of about $10^{-3}\times
\hbar k$. The momentum spectrum $\Psi_{0,q}(p)$ can be calculated
by projecting the Bloch state $|0,q\rangle$ onto plane-wave
components $|p\rangle$ (measurement basis). Using analytic
properties of the Bloch wavefunctions one obtains
\begin{equation}
\label{e.2}  \Psi_{0,q}(p)= \sqrt{\frac{2\pi}{d}} \tilde{\Phi}_0
(p)\times \sum_l\delta(p-q-2\pi l/d),~~~~~~~~~l \in \emph{}{Z}
\end{equation}
where $\tilde{\Phi}_0(p)$ is the Fourier transform of the Wannier
function of the fundamental energy band\cite{kohn}: the momentum
spectrum is composed of peaks separated by the reciprocal-lattice
vector, $2 k= 2\pi/d$, with an amplitude given by the envelope
function $\tilde{\Phi}_0(p)$. This function is not constant along
the width of the selected velocity class. Thus the center of the
measured momentum distribution is shifted with respect to the
quasi-momentum $q(t)$ (eq.$\ref{e.1}$) by a factor depending on
the momentum spread. The data analysis allowing the determination
of the Bloch period is performed in order to reduce substantially
the systematic error induced by this effect.

When we increase the interrogation time of the standing wave up to
$100~ms$, the signal is significantly degraded (loss rate becomes
larger than $50\%$). In order to understand the origin of this
losses, we have measured, for a given laser intensity, the losses
per oscillation versus the detuning $\Delta$, relative to $F=2
\rightarrow F=3$ transition (fig.~\ref{f.5}). These losses are
obtained by comparing the number of atoms measured after $N$
oscillations to those measured after $N=10$ oscillations. The
choice of the parameter $\Delta$ allows us to estimate the effects
of the spontaneous photon scattering and the interband transitions
which depend both on $\Delta$ (for interband transition see
\cite{Peik}). We observe that the losses exceed $3\%$; they are
more important than the $0.5\%$ rate losses measured using an
accelerated standing wave \cite{Battesti}. This results from the
slowness of the Bloch oscillations, since in a vertical standing
wave the Bloch period is $\sim1.2~ms$, when in reference
\cite{Battesti} the period was only $\sim 0.1~ms$.

For the experimental values of the standing wave parameters, the
losses induced by the spontaneous emission and the interband
transitions in the weak binding limit do not match with the
experimental data (dotted line in (fig.~\ref{f.5})). Performing a
least-square fit based on this model and including  collisions
with the residual Rb vapor, we extract a characteristic time
constant of the damping due to the collisions of about $70~ms$.
This value corresponds to the lifetime of the molasses for the
residual vapor pressure in the cell. Therefore we think, that the
number of Bloch oscillations, in our experiment could be increased
by reducing the pressure in the vacuum chamber.

\section{Conclusion and prospects}
We have described an experimental method to measure the vertical
velocity distribution of atoms by using a $\pi-\pi$ pulses
sequence of Doppler-sensitive Raman transitions. We have performed
a preliminary determination of the local acceleration of gravity
with a relative accuracy of $10^{-6}$ by measuring the Bloch
period. We have also demonstrated that the number of Bloch
oscillations is not yet limited by either the interband
transitions or the spontaneous emission, but only by collisions
with the background atomic vapor. To overcome this limit, we are
now building a new ultrahigh vacuum chamber where the
magneto-optical trap will be loaded by an atomic slow beam. In
order to improve the accuracy of the velocity measurement a
vibration-isolation system is also in implementation. These
improvements should allow us to take a better benefit of the Bloch
oscillations.

An attractive possibility consists to replace the $\pi-\pi$
velocity measurement by a two $\pi/2$ Ramsey-Bord\'e  sequence. We
obtain then a $\pi/2-\pi/2$-Bloch oscillation-$\pi/2-\pi/2$
atom-interferometer. Comparing this scheme to the
$\pi/2-\pi-\pi/2$ atom-interferometer used in gravimetry
\cite{Peters, Guirk}, where the pulses spacing time is limited by
the effects related to the spatial position, the advantage would
be to increase the measurement time thanks to Bloch oscillations
for a similar phase difference between the two paths. This way, we
could reduce significantly the uncertainty of the gravity
interferometric measurement. Finally, we suggest to use such
experiment nearby a Watt balance site \cite{Kibble,Eich}. In the
dynamics mode of the balance the relation which equates the
mechanical power and the electrical power is given by \cite{Eich}:
\begin{equation}
M g v = C F_J F'_J h \label{eq1}
\end{equation}
 where M is the standard mass, $v$ is the velocity of the vertical moving
coil, $h$ the Planck constant, $C$ represents a dimensionless
constant and $F_J$, $F'_J$ denote the frequencies applied to a
Josephson device. Using the $g/h$ dependence of the Bloch period
we obtain
\begin{equation}
\frac{M}{m} = \frac{C F_J F'_J }{2} \frac{\lambda}{\nu_B}
\frac{1}{v} \label{eq2}
\end{equation}
In conclusion, associating our Bloch oscillations experiment to a
Watt Balance could be used to link a macroscopic mass to an atomic
mass.

\acknowledgments We thank A.~Clairon and C.~Salomon for valuable
discussions. This experiment is supported in part by the Bureau
National de M\'etrologie (Contrats 993009 and 033006) and by the
R\'egion Ile de France (Contrat SESAME E1220).

\end{document}